\documentclass{article}
\usepackage{arxiv}
\usepackage[utf8]{inputenc} 
\usepackage[T1]{fontenc}    
\usepackage[colorlinks=true, linkcolor=blue, citecolor=blue]{hyperref}
\usepackage{url}            
\usepackage{booktabs}       
\usepackage{amsfonts}       
\usepackage{nicefrac}       
\usepackage{microtype}      
\usepackage{lipsum}
\usepackage{graphicx}
\usepackage{natbib}
\usepackage{verbatim,color,amssymb,epsfig,amsmath}
\usepackage{mathtools}
\usepackage{amsmath}
\usepackage{xcolor}
\usepackage{fancyhdr}
\usepackage{enumerate}
\usepackage{multirow}
\usepackage{array}
\usepackage[title]{appendix}
\usepackage{float}
\usepackage{dsfont}
\usepackage[ruled,linesnumbered]{algorithm2e}
\usepackage{soul}

\DeclarePairedDelimiter{\ceil}{\lceil}{\rceil}

\renewcommand{\hat}{\widehat}
\renewcommand{\tilde}{\widetilde}
\renewcommand{\bar}{\overline}

\def\boxit#1{\vbox{\hrule\hbox{\vrule\kern6pt  \vbox{\kern6pt#1\kern6pt}\kern6pt\vrule}\hrule}}

\def\boxit#1{\vbox{\hrule\hbox{\vrule\kern6pt  \vbox{\kern6pt#1\kern6pt}\kern6pt\vrule}\hrule}}

\def\one{\mathds{1}}
\def\ve{\varepsilon}
\def\CI{\mathrm{CI}}

\def\trans{^{\rm T}}
\def\half{^{1/2}}
\def\minushalf{^{-1/2}}
\def\minusone{^{-1}}

\def\minustwo{^{-2}}

\def\square{^{\,2}}

\def\IN{_{\hbox{\rm{\tiny IN}}}}

\def\E{\mathbb{E}}

\def\P{\mathbb{P}}
\def\R{\mathbb{R}}

\def\fhat{\hat{f}}

\def\ghat{\hat{g}}

\def\hbar{\bar{h}}

\def\muhat{\hat{\mu}}

\def\mutilde{\tilde{\mu}}

\def\Asc{\mathcal{A}}
\def\Bsc{\mathcal{B}}

\def\Dsc{\mathcal{D}}
\def\Fsc{\mathcal{F}}

\def\Isc{\mathcal{I}}
\def\Lsc{\mathcal{L}}

\def\Usc{\mathcal{U}}

\def\alphahat{\hat{\alpha}}

\def\gammahat{\hat{\gamma}}

\def\etahat{\hat{\eta}}
\def\thetahat{\hat{\theta}}
\def\thetahatstar{\hat{\theta}^{*}}
\def\thetatilde{\tilde{\theta}}

\def\Ehat{\hat{\E}}
\def\iton{i=1,\ldots,n}

\def\mtoM{m=1,\ldots,M}

\def\sumin{\hbox{$\sum_{i=1}^n$}}

\def\summM{\hbox{$\sum_{m=1}^M$}}

\def\DF{_{\hbox{\tiny DF}}}
\def\EX{_{\hbox{\tiny EX}}}
\def\Diag{\hbox{Diag}}

\def\hint{\hbox{$\int$}}
\def\four{^{\,4}}

\def\jzero{_{j0}}

\def\onezero{_{10}}
\def\twozero{_{20}}
\def\Tr{\mathrm{Tr}}

\def\Jsc{\mathcal{J}}

\def\betahat{\hat{\beta}}
\def\betahatmT{\betahat_m^{\,{\rm T}}}
\def\deltaZi{\delta_{Z_i}}
\def\etatilde{\tilde{\eta}}
\def\fhatYX{\fhat_{Y\mid X}}
\def\fYX{f_{Y\mid X}}
\def\Fhat{\hat{F}}
\def\FhatYX{\Fhat_{Y\mid X}}
\def\FhatYXm{\Fhat_{Y\mid X}^{\,(m)}}

\def\FYX{F_{Y\mid X}}

\def\kappahat{\hat{\kappa}}

\def\mubar{\bar{\mu}}

\def\mutilde{\tilde{\mu}}
\def\ntilde{\tilde{n}}

\def\rgtwo{r_{g,2}}
\def\rgfour{r_{g,4}}
\def\sigmahat{\hat{\sigma}}

\def\Sigmahat{\hat{\Sigma}}

\def\Ttilde{\tilde{T}}

\def\thetadot{\dot{\theta}}
\def\thetahatstar{\hat{\theta}^{\,*}}

\def\thetajzero{\theta\jzero}
\def\thetaonezero{\theta\onezero}
\def\thetaonezeroA{\thetaonezero^{(\Asc)}}
\def\thetatilde{\tilde{\theta}}
\def\thetatwozero{\theta\twozero}

\def\varrow{\overrightarrow{v}}

\def\varthetadot{\dot{\vartheta}}
\def\varthetahat{\hat{\vartheta}}
\def\xarrow{\overrightarrow{x}}
\def\Xarrow{\overrightarrow{X}}

\def\xitilde{\tilde{\xi}}

\def\ACov{\hbox{ACov}}
\def\Ehat{\hat{\E}}
\def\Normal{\hbox{Normal}}
\def\Var{\mathrm{Var}}
\def\Varhat{\hat{\Var}}
\def\suminonetonN{\hbox{$\sum_{i=n+1}^{n+N}$}}
\def\sumintilde{\hbox{$\sum_{i=1}^{\ntilde}$}}
\def\sumintilden{\hbox{$\sum_{i=\ntilde+1}^n$}}

\def\supmuB{\hbox{$\sup_{\mu\in\Bsc(\mu_0,a_n)}$}}

\def\itontilde{i=1,\ldots,\ntilde}
\def\intildeton{i=\ntilde+1,\ldots,n}
\def\inplusonetonN{i=n+1,\ldots,n+N}
\def\limninfty{\hbox{$\lim_{\,n\to\infty}$}}

\def\bse{\begin{eqnarray*}}
\def\ese{\end{eqnarray*}}
\def\be{\begin{eqnarray}}
\def\ee{\end{eqnarray}}

\newtheorem{remark}{Remark}

\newtheorem{theorem}{Theorem}

\newtheorem{proposition}{Proposition}

\newtheorem{example}{Example}

\newtheorem{assumption}{Assumption}
\newtheorem{Def}{Definition}

\title{Is External Information Useful for Data Fusion? \\ An Evaluation before Acquisition}

\author{
Guorong Dai\footnote{Guorong Dai is the first author.} \\
Department of Statistics and Data Science\\
School of Management\\ 
Fudan University\\
Shanghai, China\\
\texttt{guorongdai@fudan.edu.cn} \\
\And
Lingxuan Shao\footnote{Lingxuan Shao is the corresponding author.}\\
Department of Statistics and Data Science\\
School of Management\\ 
Fudan University\\
Shanghai, China\\
\texttt{shao\_lingxuan@fudan.edu.cn} \\
\And
Jinbo Chen\\
Department of Biostatistics\\ 
Epidemiology and Informatics, Perelman School of Medicine\\ 
University of Pennsylvania\\ 
Philadelphia, PA 19104, USA\\
\texttt{jinboche@pennmedicine.upenn.edu} \\
}

\begin{document}
\maketitle
\begin{abstract}
We consider a general statistical estimation problem involving a finite-dimensional target parameter vector. Beyond an internal data set drawn from the population distribution, external information, such as additional individual data or summary statistics, can potentially improve the estimation when incorporated via appropriate data fusion techniques. However, since acquiring external information often incurs costs, it is desirable to assess its utility beforehand using only the internal data. To address this need, we introduce a utility measure based on estimation efficiency, defined as the ratio of semiparametric efficiency bounds for estimating the target parameters with versus without incorporating the external information. It quantifies the maximum potential efficiency improvement offered by the external information, independent of specific estimation methods. To enable inference on this measure before acquiring the external information, we propose a general approach for constructing its estimators using only the internal data, adopting the efficient influence function methodology. Several concrete examples, where the target parameters and external information take various forms, are explored, demonstrating the versatility of our general framework. For each example, we construct point and interval estimators for the proposed measure and establish their asymptotic properties. Simulation studies confirm the finite-sample performance of our approach, while a real data application highlights its practical value. In scientific research and business applications, our framework significantly empowers cost-effective decision making regarding acquisition of external information.
\end{abstract}


\section{Introduction}
Consider a general statistical estimation problem, where the target is a finite-dimensional parameter vector $\mu_0\equiv\mu(F_Z)\in\R^d$, which is a functional of the distribution function $F_Z$ of a random vector $Z$. Suppose an internal data set $\Dsc\IN:=\{Z_i:\iton\}$ is available, providing $n$ independent copies of $Z$, allowing valid estimation of $\mu_0$ under suitable conditions. Additionally, external information concerning $F_Z$, independent of $\Dsc\IN$, may also be obtainable. Examples include additional individual observations or summary statistics for components of $Z$. This external information, denoted by $\Dsc\EX$, has potential to enhance the estimation of $\mu_0$ beyond what can be achieved with $\Dsc\IN$ alone, as demonstrated in \citet{chakrabortty2022semi}, \citet{hu2022paradoxes}, \citet{zhang2022high} and other data fusion literature. However, acquiring external information often incurs costs in practice. This situation raises an important question: are the potential benefits of incorporating external information worth the cost? Before making a payment decision, it is preferable to evaluate its utility for estimation using the currently available data $\Dsc\IN$. This problem is common in scientific research and business applications. To illustrate this, consider the following example, which will be explored in detail in Section \ref{sec:mean_estimation}:
\begin{example}\label{exp:mean_response}
\rm{
Suppose $Z\equiv(Y,X)\in\R\times\R^p$ consists of a response variable and a set of covariates. The goal is to estimate the mean response $\mu_0\equiv\E(Y)$. Below are three cases involving different types of external information that may be available in addition to the internal data $\Dsc\IN \equiv\{(Y_i, X_i): i = 1, \dots, 100\}$:
\begin{enumerate}[(i)]
\item An online database offers massive covariate data, almost fully capturing the true distribution $F_X$ of $X$, at a certain price.

\item Under a fixed budget, it is possible to collect either 20 additional observations of $Y$ or 100 additional observations of $X$. This situation, where response data are more expensive than covariate data, aligns with the ``semi-supervised'' setting discussed in recent literature \citep{chakrabortty2022semi, zhang2022high}.
      
\item Again under budget constraints, one may choose between 100 additional individual observations for $X$ or the average of 500 additional observations for $X$. Such scenarios are common in practice, as summary statistics are often easier to obtain than individual-level data \citep{hu2022paradoxes}.
\end{enumerate}
}
\end{example}
\noindent In Example \ref{exp:mean_response} and similar scenarios where $\mu_0$ represents other functionals of $F_Z$, it is critical to assess and compare the utility of various types of external information for estimating $\mu_0$. Motivated by this great practical importance, we aim to quantify external information utility  and perform valid statistical inference on it using the internal data $\Dsc\IN$.
  
\subsection{Basic idea}\label{sec:basic_idea}
Using the internal data $\Dsc\IN$, we can typically construct consistent and asymptotically normal estimators for common parameters, ensuring valid inference. The primary advantage of incorporating the external information $\Dsc\EX$ is its potential to enhance inference efficiency. Once $\Dsc\EX$ is obtained, it can be integrated with $\Dsc\IN$ to estimate $\mu_0$ through a variety of data fusion methods, given the rapid advancements in this field. However, it is impractical to evaluate each method and compare them to estimators relying solely on internal data. This challenge motives us to characterize the utility of $\Dsc\EX$ as its intrinsic property\textemdash independent of any specific data fusion methodology. Specifically, within a broad class of reasonable and commonly used estimators for $\mu_0$, we compare the optimal efficiency attainable in two scenarios: with and without the incorporation of $\Dsc\EX$. Such a comparison quantifies the maximum potential utility of $\Dsc\EX$ for estimating $\mu_0$, agnostic to specific algorithms employed for estimation. This conceptual framework will be formalized in Section \ref{sec:problem_formulation}.
  
\subsection{Existing relevant works and our contributions}
As one of the most compelling research areas in statistics, data fusion has garnered significant attention in the past few decades. Comprehensive reviews of this field can be found in \citet{hu2022paradoxes} and \citet{yu2024data}. Most existing studies on data fusion assume both the internal data $\Dsc\IN$ and the external information $\Dsc\EX$ are already available, concentrating on how to utilize them effectively in estimation problems. This essentially differs from our objective, which is to evaluate the intrinsic, algorithm-agnostic utility of $\Dsc\EX$ before its acquisition. To our knowledge, this has not been studied in the literature.
  
Specifically, in the context of estimating a generic functional $\mu_0\equiv\mu(F_Z)$ of the population distribution $F_Z$, we introduce in Section \ref{sec:utility_measure} a general algorithm-agnostic utility measure, denoted by $\theta_0\in(0,1\,]$, for various types of external information. Following the approach outlined in Section \ref{sec:basic_idea}, this measure is defined as the ratio of the semiparametric efficiency bounds \citep{tsiatis2007semiparametric} associated with estimation based solely on internal data versus estimation incorporating both internal data and external information. The complement $(1-\theta_0)$ quantifies the maximum potential efficiency improvement achievable by incorporating the external information $\Dsc\EX$ into the estimation. Next, we consider statistical inference for $\theta_0$ using only the internal data $\Dsc\IN\equiv\{Z_i:\iton\}$, which can be performed prior to acquiring $\Dsc\EX$. To avoid model misspecification and ensure broad applicability, we develop in Section \ref{sec:general_approach} a fully nonparametric approach for estimating $\theta_0$. The efficient influence function method \citep{hines2022demystifying} is adopted to address challenges posed by potentially complex nuisance functions, ensuring tractable asymptotic properties of our estimators. 

As illustrations of our general framework, we thoroughly analyze Example \ref{exp:mean_response} (Section \ref{sec:mean_estimation}), as well as two more complex cases where the target $\mu_0$ is a nonlinear or vector-valued functional of $F_Z$ (Appendices \ref{sec:response_quantile_individual_data}--\ref{sec:linear_regression}). In these examples with various types of external information, we specify point and interval estimators for the utility measure $\theta_0$, deriving their asymptotic properties including convergence rates and coverage guarantees. All the theoretical results are validated through comprehensive numerical studies.   Furthermore, in Section \ref{sec:discussion}, we discuss scenarios where $\Dsc\EX$ represents possibly biased information regarding $F_Z$, providing guidance on interpreting our framework in such cases.
  
\subsection{Organization}
In Section \ref{sec:problem_formulation}, we formally define the problem and introduce a utility measure for external information. The approach for constructing estimators of this measure is presented in Section \ref{sec:general_approach}. Section \ref{sec:mean_estimation} investigates a concrete example to illustrate our general framework. Simulation studies and a real data application are presented in Section \ref{sec:numerical_studies}. Finally, the article is concluded in Section \ref{sec:discussion} with several closing remarks. For another two examples, we conduct theoretical analysis in Appendices \ref{sec:response_quantile_individual_data}--\ref{sec:linear_regression}, and provide numerical results in Appendix \ref{sec:additional_numerical_results}. Necessary supplements to the main article, as well as all technical proofs, can be found in the Supplementary Material.
  
\section{Problem formulation}\label{sec:problem_formulation}
We now formalize the framework outlined in Section \ref{sec:basic_idea}. Suppose the population distribution $F_Z\in\Fsc$ for some semiparametric model $\Fsc$, and the functional $\mu$ in the definition of the target $\mu_0\equiv\mu(F_Z)\in \R^{d}$ is pathwise differentiable on $\Fsc$ at $F_Z$. As is common in standard estimation problems, we restrict attention to the class of regular and asymptotically linear estimators for $\mu_0$, which encompasses most reasonable and widely used estimators \citep{tsiatis2007semiparametric, hu2022paradoxes}. Regularity and asymptotic linearity of an estimator $\mubar$ for $\mu_0$ ensure that its limiting behaviors do not depend on the particular data generating process too heavily, and $\{\ACov(\mubar)/n\}\minushalf(\mubar-\mu_0)$ converges in distribution to a standard normal random vector as $n\to\infty$, where the asymptotic covariance matrix $\ACov(\mubar)\in\R^{d\times d}$ of $n\half\,\mubar$ is finite and non-singular. These properties underpin standard inference procedures, such as confidence interval construction and hypothesis testing, with the asymptotic efficiency of inference governed by $\ACov(\mubar)$. 

\subsection{Internal-data-only and data-fusion efficiency bounds}
  
To assess the optimal efficiency attainable with and without external information, we consider the 
corresponding semiparametric efficiency bounds, denoted by $\Theta\IN\in\R^{d\times d}$ and $\Theta\DF\in\R^{d\times d}$, for estimating $\mu_0$ based on (i) the internal data $\Dsc\IN$ alone and (ii) both $\Dsc\IN$ and the external information $\Dsc\EX$, respectively. These bounds represent meaningful lower limits (in the sense of matrix definiteness) of $\{\ACov(\mubar):\mubar\in\Usc\IN\}$ and $\{\ACov(\mubar):\mubar\in\Usc\DF\}$, where $\Usc\IN$ and $\Usc\DF$ are the sets of all regular and asymptotically linear estimators for $\mu_0$ in the respective settings. The internal-data-only efficiency bound $\Theta\IN$ is the supremum of the Cramér-Rao bounds over all parametric submodels of the semiparametric model $\Fsc$. The data-fusion efficiency bound $\Theta\DF$ can be derived analogously from a ``smaller'' semiparametric model $\Fsc\DF\subset\Fsc$, reflecting additional restrictions imposed by the external information $\Dsc\EX$. For instance, in Example \ref{exp:mean_response}(i), knowledge of the marginal distribution $F_X$ reduces the fully unrestricted nonparametric model $\Fsc$ to a subset $\Fsc\DF$, where $F_X$ is fixed and only the conditional distribution of $Y$ given $X$ remains unrestricted. Consequently, $\Fsc\DF$ contains no more parametric submodels than $\Fsc$, ensuring that $\Theta\DF$ is ``no greater than'' $\Theta\IN$, in the sense that
\begin{align}
(\Theta\IN-\Theta\DF)\hbox{ is positive semi-definite}.
\label{equ:comparison_two_bounds}
\end{align}
A crucial observation is that under given semiparametric models $\Fsc$ and $\Fsc\DF$, the efficiency bounds $\{\Theta\IN,\Theta\DF\}$ are determined solely by the population distribution $F_Z$, not depending on any specific estimation method. More formally,
\begin{align}
&\hbox{$\{\Theta\IN,\Theta\DF\}$ are functionals of $F_Z$, independent of} \nonumber \\
&\hbox{specific estimation methods used for the target $\mu_0$.}
\label{sec:property_efficiency_bounds}
\end{align}
  
We have briefly introduced the concept of the semiparametric efficiency bound in scenarios with and without external information, establishing the foundation for efficiency comparison between the two cases. For a more in-depth and comprehensive treatment of semiparametric theory in the internal-data-only setting, interested readers are referred to \citet{tsiatis2007semiparametric}, which includes methods for deriving efficiency bounds across various estimation problems. More recently, \citet{hu2022paradoxes} extended this theory to incorporate additional summary statistics. Furthermore, when the external information $\Dsc\EX$ consists of additional individual data, semiparametric theory for the combined sample structure $\Dsc\IN\cup\Dsc\EX$ (such as $\{(Y_i,X_i):\iton\}\cup\{X_i:\inplusonetonN\}$ in Example \ref{exp:mean_response}(iii)) has been developed in \citet{tsiatis2007semiparametric}. For most common parameters and various types of external information, including those considered in the examples of Section \ref{sec:mean_estimation} and Appendices \ref{sec:response_quantile_individual_data}--\ref{sec:linear_regression}, the efficiency bounds $\{\Theta\IN,\Theta\DF\}$ have already been derived in the existing literature on semiparametric theory and data fusion. Our framework enables direct applications of any established results on semiparametric efficiency bounds. Thus, we treat the forms of $\{\Theta\IN,\Theta\DF\}$ as known within our general approach, and will specify them when studying concrete examples.
  
\subsection{A utility measure for external information}\label{sec:utility_measure}
   
When the target $\mu_0$ is a vector, directly comparing the two efficiency bounds $\{\Theta\IN,\Theta\DF\}$ is unlikely to yield simple and interpretable measures. In this case, a practical strategy is to transform the matrices into scalars using the trace operation $\Tr(\cdot)$. Due to the relationship in \eqref{equ:comparison_two_bounds}, the linearity of the trace ensures
\begin{align}
\Tr(\Theta\IN)-\Tr(\Theta\DF)=\Tr(\Theta\IN-\Theta\DF)\geq 0.
\label{equ:trace_monotonicity}
\end{align}
Additionally, we assume $\Tr(\Theta\DF)>0$ to exclude the degenerate case where $\Theta\DF$ is a zero matrix, indicating $\mu_0$ can be estimated deterministically. This, along with the monotonicity in \eqref{equ:trace_monotonicity}, implies $\Tr(\Theta\IN)\geq\Tr(\Theta\DF)>0$. Then, our utility measure for external information is defined as the ratio of the two traces, denoted by $\thetaonezero$ and $\thetatwozero$, respectively:
\begin{align}
\theta_0\equiv\thetaonezero\,/\,\thetatwozero:=\Tr(\Theta\DF)\,/\,\Tr(\Theta\IN)\in(0,1\,].
\label{equ:def_theta0}
\end{align}
For the external information $\Dsc\EX$, the measure $\theta_0$ quantifies its maximum potential contribution to the estimation, as the complement $(1-\theta_0)=(\thetatwozero-\thetaonezero)/\thetatwozero\in[\,0,1)$ represents the improvement in the best achievable (componentwise) efficiency gained by incorporating $\Dsc\EX$. The extreme case of $\theta_0=1$ indicates $\Dsc\EX$ provides no additional efficiency benefit. Furthermore, as a ratio within $(0,1]$, $\theta_0$ is invariant to the scale and units of measurement, facilitating comparisons across different types of external information.
  
Besides its explicit interpretation in quantifying external information utility, the measure $\theta_0$ in \eqref{equ:def_theta0} also possesses two critical properties for our analysis, which are derived from the fact in \eqref{sec:property_efficiency_bounds}:
\begin{enumerate}[(i)]
\item It is algorithm-agnostic, defined independently of specific estimation methods used in either the internal-data-only or data-fusion setting. Formally, $\theta_0$ is a functional of $F_Z$, which represents an intrinsic characteristic of the population distribution, and thereby allows for meaningful statistical inference.
    
\item As a functional of $F_Z$, it can be estimated from the internal data $\{Z_i:\iton\}$, enabling evaluations of potential utility for external information before its acquisition.
\end{enumerate}
  
\section{General approach for constructing estimators}\label{sec:general_approach}

\paragraph{Notations.} Throughout, the lowercase letter $c$ or $c$ with a subscript, such as $c_{1}$ and $c_{2}$, stands for a generic positive constant, which may vary depending on the context. The symbol $\xrightarrow{w}$ means convergence in distribution, while $\one(\cdot)$ represents the indicator function. The notation $\|\cdot\|$ refers to the largest singular value of a matrix. For a vector $v$, let $v_{[j]}$ denote its $j$th component, and define $\overrightarrow{v}:=(1,v\trans)\trans$. The symbol $\Varhat(\cdot)$ refers to the sample variance computed from $\{Z_i:\iton\}$. For a functional $F\mapsto\vartheta(F)$, let $\varthetadot(F;\tilde{F}):=(\partial/\partial t)\vartheta(F+t\tilde{F})\mid_{t=0}$ denote its Gâteaux derivative in the direction $\tilde{F}$. Also, we use $\delta_z$ to represent the Dirac measure at the point $z$. For the standard normal distribution, let $\Phi(\cdot)$ and $u_\alpha$ denote its distribution function and $\alpha$-quantile, respectively.
        
We now turn to estimating the utility measure $\theta_0$ in \eqref{equ:def_theta0} 
using the internal data $\{Z_i:\iton\}$. To avoid model misspecification and ensure braod applicability, our analysis adopts a nonparametric framework. Since $\thetaonezero\equiv\theta_1(F_Z)$ and $\thetatwozero\equiv\theta_2(F_Z)$ may be complex functionals of $F_Z$, the resulting form of $\theta_0\equiv\theta(F_Z)$ can be intricate, often involving challenging nuisance functions. To address this complexity, we adopt the efficient influence function method. As demonstrated by \citet{hines2022demystifying}, this approach effectively mitigates the impact of nuisance estimation and, under suitable conditions, yields point estimators that achieve the optimal efficiency for $\theta_0$ in a nonparametric setting. We begin by clarifying some relevant concepts.
\begin{Def}
\label{def:nee}
Suppose a scalar-valued functional $F\mapsto\vartheta(F)$ is Gâteaux differentiable on a nonparametric model at $F_Z$, and
$\Var\{\varthetadot(F_{Z};\delta_Z-F_{Z})\}\in(0,\infty)$. In this case, $\varthetadot(F_{Z};\delta_Z-F_{Z})$ is referred to as the ``nonparametric efficient influence function'' of $\vartheta_0\equiv\vartheta(F_Z)$. If an estimator $\varthetahat$ based on $\{Z_i:\iton\}$ satisfies $\varthetahat-\vartheta_0 = n\minusone\sumin \varthetadot(F_{Z};\deltaZi-F_{Z}) + o_{p}(n\minushalf)$, then $\varthetahat$ is called a ``nonparametric efficient estimator'' for $\vartheta_0$.
\end{Def}
\noindent Our goal is to derive a nonparametric efficient estimator for $\theta_0\equiv\thetaonezero/\thetatwozero$. As indicated by the next proposition, achieving this goal reduces to constructing nonparametric efficient estimators for $\thetaonezero$ and $\thetatwozero$ individually.
  
\begin{proposition}
\label{prop:theta0}
For $\theta_0\equiv\theta(F_Z)$ defined in \eqref{equ:def_theta0}, its nonparametric efficient influence function is given by
$
\thetadot(F_{Z};\delta_Z-F_{Z})=\thetadot_1(F_{Z};\delta_Z-F_{Z})/\thetatwozero-\thetaonezero\thetadot_2(F_{Z};\delta_Z-F_{Z})/\thetatwozero^2.
$
Let $\{\thetahat_1,\thetahat_2\}$ be estimators for $\{\thetaonezero\equiv\theta_1(F_Z),\thetatwozero\equiv\theta_2(F_Z)\}$, constructed from $\{Z_i:\iton\}$. If $\Var\{\thetadot_j(F_Z;\delta_Z-F_Z)\}<\infty$, and $\thetahat_j-\thetajzero=n\minusone\sumin\thetadot_j(F_{Z};\deltaZi-F_{Z})+O_p(r_j)$ for some positive series $r_j\equiv r_j(n)\to 0$ ($j=1,2$), then the ratio estimator $\thetahat:=\thetahat_1/\thetahat_2$ satisfies
\begin{align*}
\thetahat-\theta_0=n\minusone\sumin\thetadot(F_{Z};\deltaZi-F_{Z})+O_p(r_1+r_2+n\minusone).
\end{align*}
\end{proposition}
  
To estimate $\thetajzero\equiv\theta_j(F_Z)$, we apply the estimation equation method from \citet{hines2022demystifying}, solving $n\minusone\sumin\thetadot_j(\Fhat_Z;\deltaZi-\Fhat_Z)=0$ to obtain an estimator $\thetahat_j\equiv\theta_j(\Fhat_Z)$ ($j=1,2$), where $\Fhat_Z$ is an estimator for $F_Z$. This procedure is computationally straightforward because $\thetajzero$ is defined as the trace of an efficiency bound, corresponding to a covariance matrix. This implies $\thetajzero\equiv\E\{\psi_j(Z;F_Z)\}\equiv\int\psi_j(z;F_Z)dF_Z(z)$ for some functional $\psi_j(\cdot\,;\cdot)$. Consequently, its Gâteaux derivative has the form
\begin{align*}
\thetadot_{j}(F_{Z};\delta_Z-F_{Z})=\phi_j(Z;F_Z)+\psi_j(Z;F_Z)-\theta_j(F_Z),
\end{align*}
with $\phi_j(Z;F_Z):=(\partial/\partial t)\{ \hint\psi_{j}(z;F_Z+t(\delta_Z-F_Z))dF_Z(z)\} \mid_{t=0}$. This leads to the closed-form expression of $\thetahat_j$:
\begin{align*}
\thetahat_j\equiv\theta_j(\Fhat_Z)=n\minusone\sumin\{\phi_j(Z_i;\Fhat_Z)+\psi_j(Z_i;\Fhat_Z)\}.
\end{align*}
In concrete examples such as those in Section \ref{sec:mean_estimation} and Appendices \ref{sec:response_quantile_individual_data}--\ref{sec:linear_regression}, we allow $\Fhat_Z$ to be any reasonable estimator for $F_Z$, such that 
\begin{enumerate}[(i)]
\item the marginal distributions of subvectors of $Z$ are approximated by their empirical counterparts from $\{Z_i:\iton\}$;
    
\item characteristics of the conditional distributions, such as means and distribution functions, are estimated using flexible regression or machine learning methods that satisfy certain high-level convergence rate conditions;
    
\item the marginal and conditional density functions are estimated using kernel smoothing.
\end{enumerate}
  
We summarize the steps for constructing a point estimator of $\theta_0$ as follows: 
\vspace{0.25cm}
  
\begin{algorithm}[H]
\caption{Construction of a point estimator for $\theta_0$.}
\label{algo:point_estimator}
Compute the Gâteaux derivative $\thetadot_j(F_{Z};\delta_Z-F_{Z})$ for $j=1,2$. 
    
Solve $n\minusone\sumin\thetadot_j(\Fhat_Z;\deltaZi-\Fhat_Z)=0$ to obtain an estimator $\thetahat_j$ for $\thetajzero$ ($j=1,2$). 
    
Take the ratio $\thetahat\equiv\thetahat_1\,/\,\thetahat_2$ as the final estimator for $\theta_0$ in \eqref{equ:def_theta0}.
\end{algorithm}
\vspace{0.25cm}
\noindent In various concrete examples presented in Sections \ref{sec:mean_estimation} and Appendices \ref{sec:response_quantile_individual_data}--\ref{sec:linear_regression}, we will apply Algorithm \ref{algo:point_estimator} and demonstrate that the resulting point estimators achieve nonparametric efficiency under suitable conditions.
  
\section{Example: Mean response estimation with external covariate information}\label{sec:mean_estimation}
To illustrate our general framework, we analyze a concrete example: mean response estimation with external covariate information, where $Z\equiv(Y,X)\in\R\times\R^p$, and the objective is to estimate $\mu_0\equiv\E(Y)$ within a nonparametric model. Using only the internal data $\{(Y_i,X_i):\iton\}$, the efficiency bound for $\mu_0$ is given by
\begin{align}
\thetatwozero\equiv\E\{(Y-\mu_0)^2\},
\label{equ:thetatwozero_response_mean}
\end{align}
as shown by \citet{tsiatis2007semiparametric}. We assess the utility of two types of external covariate information: individual data or a summary statistic, that is,
\begin{align}
&\hbox{$N$ independent copies $\{X_i:\inplusonetonN\}$ of $X$, or } \label{equ:individual_data}\\
&\hbox{their average $N\minusone\suminonetonN X_i$}.
\label{equ:summary_statistic}
\end{align}
Define the fixed constant $\nu:=n/(n+N)\in[\,0,1)$, which is known prior to acquiring the external information. When $\nu=0$, corresponding to $N=\infty$, the external information in \eqref{equ:individual_data} and \eqref{equ:summary_statistic} represents the distribution and expectation of $X$, respectively. Incorporating this information, \citet{zhang2022high} and \citet{hu2022paradoxes} demonstrated that for estimating $\mu_0$, the data-fusion efficiency bound is given by
\begin{align}
\thetaonezero\equiv\E[(1-\nu)\{Y-g(X)\}^2+\nu(Y-\mu_0)^2],
\label{equ:thetaonezero_response_mean}
\end{align}
where $g(X)$ equals $\E(Y\mid X)$ in case \eqref{equ:individual_data} and $\beta_0\trans\Xarrow$ in case \eqref{equ:summary_statistic}, with $\beta_0:=\E(\Xarrow\Xarrow\trans)\minusone \E(\Xarrow Y)$.
We proceed with the general formula \eqref{equ:thetaonezero_response_mean} of $\thetaonezero$, where $g(X)\in\{\E(Y\mid X),\beta_0\trans \Xarrow\}$, and analyze the two cases, \eqref{equ:individual_data} and \eqref{equ:summary_statistic}, simultaneously. Combining \eqref{equ:thetatwozero_response_mean} and \eqref{equ:thetaonezero_response_mean}, the utility measure $\theta_0$ defined in \eqref{equ:def_theta0} takes the form
\begin{align}
\theta_0\equiv (1-\nu)\E[\{Y-g(X)\}^2]\,\big/\,\E\{(Y-\mu_0)^2\}+\nu.
\label{equ:thetazero_response_mean}
\end{align}
To apply Algorithm \ref{algo:point_estimator}, we compute the nonparametric efficient influence functions of $\{\thetaonezero,\thetatwozero,\theta_0\}$, as detailed in the next proposition. This provides a road map for constructing point estimators.
\begin{proposition}
\label{prop:EIF_response_mean}
The nonparametric efficient influence functions of $\thetaonezero\equiv\theta_1(F_{Z})$ in \eqref{equ:thetaonezero_response_mean}, $\thetatwozero\equiv\theta_2(F_{Z})$ in \eqref{equ:thetatwozero_response_mean} and $\theta_{0}\equiv \thetaonezero/\thetatwozero$ in \eqref{equ:thetazero_response_mean} are given as follows:
\begin{align*}
&\thetadot_1(F_{Z};\delta_Z-F_{Z})=(1-\nu)\{Y-g(X)\}^2+\nu(Y-\mu_0)^2-\thetaonezero, \\
&\thetadot_2(F_{Z};\delta_Z-F_{Z})=(Y-\mu_0)^2-\thetatwozero,\\
&\thetadot(F_{Z};\delta_Z-F_{Z})=\thetadot_1(F_{Z};\delta_Z-F_{Z})/\thetatwozero-\thetaonezero\thetadot_2(F_{Z};\delta_Z-F_{Z})/\thetatwozero^2.
\end{align*}
\end{proposition} 
  
\subsection{Point estimation}
Using the results in Proposition \ref{prop:EIF_response_mean}, we proceed to step 2 of Algorithm \ref{algo:point_estimator} and derive the following point estimators for $\{\thetaonezero,\thetatwozero\}$ based on the internal data $\{(Y_i,X_i):\iton\}$:
\begin{align}
\thetahat_1\equiv n\minusone\sumin[(1-\nu)\{Y_i-\ghat(X_i)\}^2+\nu(Y_i-\muhat)^2], \quad \thetahat_2\equiv n\minusone\sumin(Y_i-\muhat)^2,
\label{equ:thetahatone_thetahattwo_mean_response}
\end{align}
where $\muhat:=n\minusone\sumin Y_i$. To construct $\{\ghat(X_i):\iton\}$, we adopt the cross-fitting strategy \citep{chernozhukov2018double}. Specifically, for a fixed integer $M\geq 2$, we randomly divide the index set $\Isc:=\{1,\ldots,n\}$ into $M$ disjoint subsets $\{\Isc_1,\ldots,\Isc_M\}$ of sizes as even as possible, and set
\begin{align}
\ghat(X_i)\equiv\summM\ghat_m(X_i)\one(i\in\Isc_m),
\label{equ:cf_example1}
\end{align}
where $\ghat_m$ is an estimator for $g$ constructed using the data $\{(Y_i,X_i):i\in\Isc\backslash\Isc_m\}$. The cross-fitting procedure removes the dependence between $\ghat$ and $X_i$ in $\ghat(X_i)$ for $\iton$. It simplifies the control of remainder terms in the expansion of $\thetahat_1$, while not altering the influence function thereof. This approach allows for the use of flexible regression or machine learning methods, with potentially intractable first-order properties, to construct $\ghat_m$. Therefore, we leave the form of $\ghat_m$ unspecified and only impose high-level convergence rate conditions; see Assumption \ref{ass:response_mean}. In the numerical studies presented in Section \ref{sec:numerical_studies}, various methods will be employed to construct $\ghat_m$. Finally, the point estimator for $\theta_0$ in \eqref{equ:thetazero_response_mean} is given by
\begin{align}
\thetahat\equiv\thetahat_1/\thetahat_2=(1-\nu)\sumin\{Y_i-\ghat(X_i)\}^2/\{\sumin(Y_i-\muhat)^2\}+\nu.
\label{equ:thetahat_response_mean}
\end{align}
In the following, we present asymptotic properties of $\thetahat$ after specifying necessary regularity conditions.
\begin{assumption}\label{ass:response_mean}
\rm{
For $\mtoM$ and some positive series $\rgtwo,\rgfour\to 0$ as $n\to\infty$, the estimator $\ghat_m$ in \eqref{equ:cf_example1} satisfies
\begin{align*}
\hint\{\ghat_m(x)-g(x)\}^{2} dF_X(x)=O_p(\rgtwo^2),\ \hint\{\ghat_m(x)-g(x)\}^{4} dF_X(x)=O_p(\rgfour^4).
\end{align*}
}
\end{assumption}
  
\begin{theorem}
\label{thm:exm1}
Suppose Assumption~\ref{ass:response_mean} holds and there exists some constant $c>0$ such that $\E[\{Y-g(X)\}^2\mid X]+\E(Y^4)<c$ almost surely.
Then, our estimator $\thetahat$ in \eqref{equ:thetahat_response_mean} satisfies
\begin{align}
\thetahat-\theta_0
= n\minusone\sumin\thetadot(F_{Z};\deltaZi-F_{Z}) + O_p(\rgtwo^2 + (\rgtwo+\rgfour^{2})/n\half + n\minusone),
\label{equ:expansion_response_mean}
\end{align}
with $\thetadot(F_{Z};\deltaZi-F_{Z})$ specified in Proposition \ref{prop:EIF_response_mean}. It follows that:  
\begin{enumerate}[(a)]
      
\item $\thetahat$ attains the following convergence rate:
\begin{align}
\thetahat-\theta_0
= O_p(n\minushalf + \rgtwo^2 + (\rgtwo+\rgfour^{2})/n\half + n\minusone)=O_p(n\minushalf + \rgtwo^2); \label{equ:rate_mean_response}
\end{align}
      
\item if $\rgtwo^2 + (\rgtwo+\rgfour^{2})/n\half=o(n\minushalf)$ and $\P\{\thetadot(F_{Z};\delta_Z-F_{Z})= 0\}<1$, then $\thetahat$ is nonparametric efficient, that is, $\thetahat-\theta_0=n\minusone\sumin\thetadot(F_{Z};\deltaZi-F_{Z})+o_p(n\minushalf)$;
      
\item if $\P\{g(X)=\mu_0\}=1$, then $\P\{\thetadot(F_{Z};\delta_Z-F_{Z})= 0\}=1$ and the rate in \eqref{equ:rate_mean_response} is improved to 
\begin{align}
\thetahat-\theta_0=O_p(\rgtwo^2 +(\rgtwo+\rgfour^{2})/n\half + n\minusone).
\label{equ:improved_rate_mean_response}
\end{align}
      
\end{enumerate}
\end{theorem}
  
\begin{remark}[Discussion of the rate conditions on $\ghat_m$]
\rm{
Theorem \ref{thm:exm1}(b) requires the following conditions:
\begin{align}
&\hint\{\ghat_m(x)-g(x)\}^2dF_X(x)=o_p(n\minushalf), \label{equ:rgtwo} \\
&\hint\{\ghat_m(x)-g(x)\}^4dF_X(x)=o_p(1).  \label{equ:rgfour}
\end{align}
It is evident that under \eqref{equ:rgtwo}, condition \eqref{equ:rgfour} holds, for instance, as long as $\{\ghat_m(X),\,g(X)\}$ are almost surely bounded. We now address condition \eqref{equ:rgtwo} in the two cases separately:
\begin{enumerate}[(a)]
\item In case \eqref{equ:individual_data} with $g(X)\equiv\E(Y\mid X)$, the estimator $\ghat_m$ can be constructed using various regression or machine learning algorithms targeting the conditional mean $\E(Y\mid X)$. In the context of estimation problems involving nuisance functions, such as those in \citet{chernozhukov2018double}, the high-level condition \eqref{equ:rgtwo} is often imposed as the general minimum convergence rate requirement for flexible nuisance estimators with unspecified forms.
      
\item In case \eqref{equ:summary_statistic} with $g(X)\equiv\beta_0\trans \Xarrow$, we set the nuisance estimator $\ghat_m(X_i)\equiv\betahatmT \,\Xarrow_i$ in \eqref{equ:cf_example1}, where $\betahat_m$ is an estimator for $\beta_0\equiv\E(\Xarrow\Xarrow\trans)\minusone \E(\Xarrow Y)$, constructed using the data $\{(Y_i,X_i):i\in\Isc\backslash\Isc_m\}$. For this setting, condition \eqref{equ:rgtwo} can be rewritten as
\begin{align*}
\hint(\betahatmT\,\xarrow-\beta_0\trans\, \xarrow)^2dF_X(x)=(\betahat_m-\beta_0)\trans\E(\Xarrow\Xarrow\trans)(\betahat_m-\beta_0)=o_p(n\minushalf).
\end{align*}
Therefore, if $\|\E(\Xarrow\Xarrow\trans)\|<c$ for some constant $c>0$, a sufficient condition for \eqref{equ:rgtwo} is
\begin{align}
\|\betahat_m-\beta_0\|^2=o_p(n\minushalf).
\label{equ:rate_betahatm}
\end{align}
When the dimension $p$ of $X$ is fixed, \eqref{equ:rate_betahatm} is satisfied by the ordinary least squares estimator under standard conditions. In high-dimensional settings with a diverging number of covariates in $X$, a reasonable choice for $\betahat_m$ is the Lasso estimator. For such $\betahat_m$, \eqref{equ:rate_betahatm} is typically equivalent to a sparsity assumption\textemdash the number of nonzero components in the vector $\beta_0$ is of order $o(n\half/\log p)$ \citep{wainwright2019high}.
\end{enumerate}
}
\end{remark}
  
\begin{remark}[Interpretation of Theorem \ref{thm:exm1}(c)]\label{remark:interpretation_improved_rate}
\rm{
The condition $\P\{g(X)=\mu_0\}=1$ is equivalent to the utility measure $\theta_0\equiv (1-\nu)\E[\{Y-g(X)\}^2]\,\big/\,\E\{(Y-\mu_0)^2\}+\nu=1$. This implies incorporating the external information does not provide any efficiency benefit for estimating $\mu_0\equiv\E(Y)$. In this scenario, the influence function $\thetadot(F_Z;\delta_Z-F_Z)$ in the expansion \eqref{equ:expansion_response_mean} vanishes, leading to the improved rate in \eqref{equ:improved_rate_mean_response}.
}
\end{remark}

\subsection{Confidence interval construction}\label{sec:CI_mean}
Theorem \ref{thm:exm1}(c) indicates if $\P\{g(X)=\mu_0\}=1$, meaning $\theta_0=1$, the first-order term in the expansion \eqref{equ:expansion_response_mean}, $n\minusone\sumin\thetadot(F_{Z};\deltaZi-F_{Z})$, is almost surely zero. In this case, the limiting behavior of $(\thetahat-\theta_0)$ is governed by the higher-order terms, which include errors from estimating the nuisance function $g$. Consequently, the asymptotic distribution of $\thetahat$ differs depending on whether $\theta_0\in(0,1)$ or $\theta_0=1$. Additionally, when $g(X)$ represents a nonparametric or high-dimensional model, these higher-order terms often exhibit intractable limiting behaviors. In practice, we generally do not know if $\P\{g(X)=\mu_0\}=1$, preventing $\thetahat$ from being used to construct confidence intervals that are valid for all $\theta_0\in(0,1]$. This issue motivates us to propose an alternative estimator with a consistently well-behaved asymptotic distribution.
  
The term $n\minusone\sumin\thetadot(F_{Z};\deltaZi-F_{Z})$ in \eqref{equ:expansion_response_mean} corresponds to the errors in approximating the expectations $\E[\{Y-g(X)\}^2]$ and $\E\{(Y-\mu_0)^2\}$ (as if $g$ and $\mu_0$ were known), as demonstrated in the Supplementary Material. Specifically, these errors are given by
\begin{align}
n\minusone\sumin\{Y_i-g(X_i)\}^2-\E[\{Y-g(X)\}^2]~\hbox{and}~n\minusone\sumin(Y_i-\mu_0)^2-\E\{(Y-\mu_0)^2\}.
\label{equ:error_terms_thetahat}
\end{align}
Whenever $\P\{g(X)=\mu_0\}=1$, the two terms in \eqref{equ:error_terms_thetahat} are almost surely identical and cancel out, resulting in $\P\{n\minusone\sumin\thetadot(F_{Z};\deltaZi-F_{Z})=0\}=1$. To avoid this vanishing first-order term issue, our strategy is to approximate the two expectations using different halves of the available data. Specifically, we propose an alternative estimator:
\begin{align}
\textstyle \thetatilde\equiv(1-\nu) \sumintilde\{Y_i-\ghat(X_i)\}^2\,\big/\,\{\sumintilden(Y_i-\muhat)^2\}+\nu~\hbox{with}~\ntilde:=\ceil{n/2},
\label{equ:def_thetatilde}
\end{align}
where $\{\ghat(X_i):\itontilde\}$ are constructed as in \eqref{equ:cf_example1}, and $\muhat\equiv n\minusone\sumin Y_i$. The error terms of $\thetatilde$, corresponding to those in \eqref{equ:error_terms_thetahat}, are now
\begin{align*}
\ntilde\minusone\sumintilde\{Y_i-g(X_i)\}^2-\E[\{Y-g(X)\}^2]~\hbox{and}~(n-\ntilde)\minusone\sumintilden(Y_i-\mu_0)^2-\E\{(Y-\mu_0)^2\},
\end{align*}
which are unequal and do not cancel out, regardless of whether $\P\{g(X)=\mu_0\}=1$. As a result, under suitable conditions, we can ensure these error terms dominate the higher-order terms in the expansion of $(\thetatilde-\theta_0)$. This allows for an explicit asymptotic distribution of $\thetatilde$, enabling the construction of valid confidence intervals for $\theta_0$.
  
\begin{theorem}
\label{thm:exm1:inference}
Suppose Assumption \ref{ass:response_mean} holds with $n\half\rgtwo^2+\rgfour=o(1)$, and there exist some positive constants $\{c_1,c_2\}$ such that $\E[\{Y-g(X)\}^2\mid X]+\E(|Y|^{4+c_1})<c_2$ almost surely. If 
\begin{align}
\gamma^2:= 2(1-\nu)^2\,\Var[\{Y-g(X)\}^2]\,/\,\thetatwozero\square+2(\thetaonezero-\nu\thetatwozero)^2\,\Var\{(Y-\mu_{0})^2\}\,/\,\thetatwozero\four >c
\label{equ:variance_mean_response}
\end{align}
for some constant $c>0$, then we have the following conclusions:
\begin{enumerate}[(a)]
\item the estimator $\thetatilde$ in \eqref{equ:def_thetatilde} satisfies $n\half(\thetatilde-\theta_0)/\gamma\,\xrightarrow{w}\,\Normal(0,1)$;
      
\item for any predetermined confidence level $\alpha\in(0,1)$, the interval
\begin{align}
\CI_\alpha\!:=\![\,\thetatilde-\gammahat\,u_{(1+\alpha)/2}/n\half,\, \thetatilde+\gammahat\,u_{(1+\alpha)/2}/n\half\,]\hbox{ satisfies }\limninfty\,\P(\theta_0\in\CI_\alpha)=\alpha,
\label{equ:CI_mean_response}
\end{align}
where
$
\gammahat\square:=2(1-\nu)^2\,\Varhat[\{Y-\ghat(X)\}^2]\,/\,\thetahat_{2}\square+2(\thetahat_{1}-\nu\thetahat_{2})^2\,\Varhat\{(Y-\muhat)^2\}\,/\,\thetahat_{2}\four,
$
with $\{\thetahat_1,\thetahat_2\}$ as in \eqref{equ:thetahatone_thetahattwo_mean_response}, $\{\ghat(X_i):\iton\}$ as in \eqref{equ:cf_example1} and $\muhat\equiv n\minusone\sumin Y_i$.
\end{enumerate}
\end{theorem}
  
\noindent As indicated by \eqref{equ:variance_mean_response}, as long as one of the two variances, $\Var[\{Y-g(X)\}^2]$ and $\Var\{(Y-\mu_{0})^2\}$, is positive, the asymptotic variance $\gamma^2$ of $\thetatilde$ has a positive lower bound. This ensures the asymptotic normality result stated in Theorem \ref{thm:exm1:inference}(a). Moreover, with a consistent estimate $\gammahat$ for $\gamma$, we provide in Theorem \ref{thm:exm1:inference}(b) an asymptotically valid confidence interval $\CI_\alpha$ for $\theta_0$.

\section{Numerical studies}\label{sec:numerical_studies}
In this section, we evaluate the numerical performance of the method introduced in Section \ref{sec:mean_estimation}. When the nuisance function $g(X)\equiv\E(Y\mid X)$, we estimate it using the Super Learner algorithm \citep{van2007super}, which combines estimators from ordinary least squares regression, additive models \citep{hastie1990generalized} and neural networks \citep{ripley2007pattern} into an optimal convex combination. The weights for the combination are determined by minimizing the squared prediction error using ten-fold cross-validation. These algorithms are implemented using the \texttt{R} packages \texttt{SuperLearner}, \texttt{glm}, \texttt{gam} and \texttt{nnet} with all arguments set to the default values. For the case where $g(X)\equiv\beta_0\trans\Xarrow$, we approximate the parameter vector $\beta_0\equiv\E(\Xarrow\Xarrow\trans)\minusone\E(\Xarrow Y)$ using the ordinary least squares estimate. In the cross-fitting procedure \eqref{equ:cf_example1}, we set the number of folds to $M=5$.
  
Since the utility measure $\theta_0$ lies within the interval $(0,1]$, it is practical to truncate its estimators. For a general interval $\Lsc := [\ell_1, \ell_2]$ with $\ell_1 \leq \ell_2$, which reduces to a single value if $\ell_1 = \ell_2$, we define the truncating function as
\begin{align}
H(\Lsc):=[\,\ell_1\one(0< \ell_1< 1)+\one(\ell_1\geq 1),\,\ell_2\one(0< \ell_2<1)+\one(\ell_2\geq 1)\,].
\label{equ:truncation}
\end{align}
We apply this truncating function to the point estimator $\thetahat$ and confidence interval $\CI_\alpha$, which are computed as in  \eqref{equ:thetahat_response_mean} and \eqref{equ:CI_mean_response}. In all numerical studies, when referring to $\thetahat$ and $\CI_\alpha$, we assume they have been truncated by the function $H(\cdot)$. 
  
\subsection{Simulations}
\label{sec:simu}
We draw the random vector $X\equiv(S,W)\in\R^2$ from the bivariate normal distribution with $\E(S)=\E(W)=0$, $\E(S^2)=\E(W^2)=1$ and $\rho:=\E(SW)=0.2$. The response variable $Y$ is generated by the model 
\begin{align}
Y=b(S+W)+\varepsilon \,\hbox{ with }\, b\in\{0,0.5,1.0\},
\label{equ:simulation_model}
\end{align}
where $\ve$ is a standard normal variable independent of $X$. In this setting, the true value of the utility measure $\theta_0$ is given by $\theta_0=(1-\nu)/\{2b^2(1+\rho)+1\}+\nu$ according to the formula \eqref{equ:thetazero_response_mean}, where we set the coefficient $\nu=1/2$. All results are summarized over $1000$ iterations.
  
For the utility measure $\theta_0$, we construct the point estimator $\thetahat$ and $95\%$ confidence interval $\CI_{0.95}$ using internal data of size $n\in\{500,1000,2000\}$, as outlined in \eqref{equ:thetahat_response_mean} and \eqref{equ:CI_mean_response}. Table \ref{tab:simulation_mean_response} displays the means and standard deviations of the absolute error $|\thetahat-\theta_0|$, along with the average lengths and coverage rates of $\CI_{0.95}$. The results clearly show the estimation errors of $\thetahat$ are negligible in all cases. We also notice that when $b=0$, which implies $\theta_0=1$, the estimation errors converge substantially faster than in the other scenarios. This observation is consistent with the discussion in Remark \ref{remark:interpretation_improved_rate}: our point estimator $\thetahat$ achieves the improved convergence rate in \eqref{equ:improved_rate_mean_response} if $\theta_0=1$. As for the confidence intervals, they exhibit satisfactory coverage rates, nearly matching the nominal level $95\%$ when $b\in\{0.5,1.0\}$. The over-coverage in cases with $b=0$, where $\theta_0=1$, is a result of the truncation procedure defined in \eqref{equ:truncation}. Specifically, when $\theta_0=1$, which lies at the boundary of the parameter space $(0,1]$, intervals such as $[1.1,1.2]$, with lower bounds greater than one and thus unable to cover $\theta_0$, are truncated to $\{1\}$ by the function $H(\cdot)$. This truncation operation enhances the coverage rates of confidence intervals without increasing their lengths. Overall, the results for confidence intervals validate the conclusions of Theorem \ref{thm:exm1:inference} and demonstrate the ability of our method to perform valid inference for external information utility.

\begin{table}
\begin{center}
\caption{Simulation results for the point estimator $\thetahat$ in \eqref{equ:thetahat_response_mean} and $95\%$ confidence interval $\CI_{0.95}$ in \eqref{equ:CI_mean_response}: Means (MAE) and standard deviations (SDAE) of the absolute error $|\thetahat-\theta_0|$, as well as the average lengths (AL) and coverage rates (CR) of $\CI_{0.95}$. Here $b$ is the coefficient in model \eqref{equ:simulation_model}, $n$ is the sample size of internal data, $g$ is the nuisance function in the formula \eqref{equ:thetazero_response_mean} of $\theta_0$, and $\beta_0\equiv\E(\protect\Xarrow\protect\Xarrow\trans)\minusone\E(\protect \Xarrow Y)$. \ul{The numbers in all columns, except for the first and second, have been multiplied by $100$}. \label{tab:simulation_mean_response} \vspace{3mm}}
\renewcommand{\arraystretch}{1}
\begin{tabular}{|cc|cccc|cccc|}
\hline
&      & \multicolumn{4}{c|}{$g(X)\equiv\E(Y\mid X)$} & \multicolumn{4}{c|}{$g(X)\equiv\beta_0\trans\Xarrow$} \\
$b$                  & $n$  & MAE      & SDAE     & AL      & CR    & MAE       & SDAE      & AL       & CR     \\
\hline
\multirow{3}{*}{$0$}   & 500  & 0.0115 & 0.0755 & 15.6998 & 99.5 & 0.0109 & 0.0630 & 14.5508 & 99.3 \\
& 1000 & 0.0046 & 0.0362 & 10.7432 & 98.7 & 0.0057 & 0.0305 & 10.4498 & 98.4 \\
& 2000 & 0.0038 & 0.0212 & 7.9152  & 98.2 & 0.0048 & 0.0226 & 7.3575  & 98.0 \\
\hline
\multirow{3}{*}{$0.5$} & 500  & 1.4330 & 1.0899 & 15.6894 & 95.1 & 1.3960 & 1.0451 & 15.6162 & 94.9 \\
& 1000 & 0.9868 & 0.7457 & 11.0005 & 93.6 & 0.9784 & 0.7349 & 10.9841 & 93.6 \\
& 2000 & 0.6669 & 0.5127 & 7.7823  & 94.2 & 0.6630 & 0.5085 & 7.7749  & 94.4 \\
\hline
\multirow{3}{*}{$1.0$} & 500  & 0.8990 & 0.7084 & 7.4180  & 94.8 & 0.8815 & 0.6785 & 7.3626  & 95.0 \\
& 1000 & 0.6341 & 0.4811 & 5.1764  & 94.4 & 0.6306 & 0.4781 & 5.1698  & 94.3 \\
& 2000 & 0.4349 & 0.3343 & 3.6622  & 94.0 & 0.4339 & 0.3330 & 3.6595  & 93.9  \\
\hline
\end{tabular}
\end{center}
\end{table}

\subsection{A real data example}
We now apply our the method from Section \ref{sec:mean_estimation} to data from the National Health and Nutrition Examination Survey (\url{https://wwwn.cdc.gov/nchs/nhanes/continuousnhanes/default.aspx?BeginYear=2015}), including records of systolic blood pressure ($Y\in\R$), age and body mass index ($X\in\R^2$) for $n=6550$ individuals. These internal data allow for estimating the population's average systolic blood pressure, $\E(Y)$, a key public health metric. To improve estimation precision, the most effective approach is to collect additional observations for $Y$. However, this can be costly, as measuring blood pressure often requires multiple readings using specialized instruments over an extended period. In contrast, individual data and summary statistics of $X$, though less effective, are easier to obtain from demographic or healthcare databases. Therefore, we consider three types of external information: 
\begin{align}
\hbox{(i) $\{Y_i:\inplusonetonN\}$, (ii) $\{X_i:\inplusonetonN\}$, (iii) $N\minusone\suminonetonN X_i$,} 
\label{equ:three_types_real_data}
\end{align}
and evaluate their utility for estimating $\E(Y)$ using the internal data $\{(Y_i,X_i):\iton\}$. Comparing their effectiveness with acquisition costs helps determine the most worthwhile option.
  
The utility of (i)--(iii) in \eqref{equ:three_types_real_data} is quantified by $(1-\theta_0)$, with $\theta_0$ defined in \eqref{equ:def_theta0}. This parameter represents the potential efficiency gain from incorporating external information. The utility of (i) is $1-\theta_0\equiv 1-\nu\equiv 1-n/(n+N)$, serving as a benchmark. For (ii) and (iii), we focus on $\theta_0^*:=(1-\theta_0)/(1-\nu)\equiv 1-\E[\{Y-g(X)\}^2]\,/\,\E\{(Y-\mu_0)^2\}$, with $\theta_0$ specified in \eqref{equ:thetazero_response_mean}, which measures their relative utility compared to (i). Since $\theta_0^*$ is a linear transformation of $\theta_0$, its point and interval estimators follow from $\thetahat$ and $\CI_\alpha\equiv[L_\alpha,U_\alpha]$ in \eqref{equ:thetahat_response_mean} and \eqref{equ:CI_mean_response}, given by $\thetahatstar:=(1-\thetahat)/(1-\nu)$ and $\CI_\alpha^*:=[\,(1-U_\alpha)/(1-\nu),\,(1-L_\alpha)/(1-\nu)\,]$. Applying the inference method from Section \ref{sec:mean_estimation}, we obtain: 
\begin{itemize}
\item for (ii), $\thetahatstar=0.2864$ and $\CI_{0.95}^*=[\,0.2128,0.3288\,]$;
    
\item for (iii), $\thetahatstar=0.2580$ and $\CI_{0.95}^*=[\,0.1799,0.3001\,]$.
\end{itemize}
The close point estimates and overlapping confidence intervals indicate that the two types of external information offer similar utility. This suggests the conditional mean model $\E(Y\mid X)$ is likely close to linear. Given the summary statistic (iii) is typically easier to obtain than the individual data (ii), (iii) is the more cost-effective choice. Between (i) and (iii), the decision depends on their respective costs and specific budget constraints.

\section{Discussion}\label{sec:discussion}
\subsection{Additional examples}
For external information $\Dsc\EX$ in various forms, we developed a general framework to assess its utility for estimating a generic functional $\mu_0\equiv\mu(F_0)$ of the population distribution $F_Z$ prior to its acquisition. Beyond the practically important example in Section \ref{sec:mean_estimation}, Appendices \ref{sec:response_quantile_individual_data}--\ref{sec:linear_regression} examine two additional cases that are both practically relevant and technically more involved. There, $\mu_0$ corresponds to either a response quantile or the parameter vector of a multiple linear model, making it a nonlinear or vector-valued functional of $F_Z$, while $\Dsc\EX$ consists of either individual covariate data or univariate regression parameter estimates. For these settings, we specify the utility measure $\theta_0$ in \eqref{equ:def_theta0} and establish theoretical properties for its point and interval estimators, constructed using the approach from Section \ref{sec:general_approach}. The corresponding numerical studies appear in Appendix \ref{sec:additional_numerical_results}. These examples further demonstrate the versatility of our framework.
  
\subsection{Possibly biased external information}
While our primary focus has been on cases where the external information $\Dsc\EX$ accurately reflects characteristics of the population distribution $F_Z$, the utility measure $\theta_0$ in \eqref{equ:def_theta0} also has meaningful interpretations when $\Dsc\EX$ represents possibly biased information concerning $F_Z$. To illustrate this, recall the mean response estimation problem from Section \ref{sec:mean_estimation}, where the target parameter is $\E(Y)$, and the internal data set $\{(Y_i,X_i):\iton\}$ consists of $n$ independent copies of the pair $(Y,X)\in\R\times\R^p$. The external information is now $\xitilde:=N\minusone\suminonetonN (X_i+h)$, where $h\in\R^p$ is an unobserved constant vector. Suppose there exists an unknown index set $\Asc\subset\Jsc:=\{1,\ldots,p\}$ such that $h_{[j]}=0$ for $j\in\Asc$ and $h_{[j]}\neq 0$ for $j\in\Jsc\backslash\Asc$. This indicates $\xitilde$ is a partially biased estimate for $\E(X)$. In this scenario, directly integrating $\xitilde$ may lead to biased estimation. As shown in \citet{hu2022paradoxes}, the oracle case occurs when $\Asc$ is known, allowing bias to be avoided by integrating only the components in $\xitilde_\Asc$\footnote{For a vector $v\in\R^p$, we let $v_\Asc$ denote its subvector consisting of the components indexed by $\Asc$, and define $\varrow_{\!\!\!\Asc}:=(1,v_{\!\Asc}\trans)\trans$ with the convention $\varrow_{\!\!\emptyset}:=1$.}. The corresponding data-fusion efficiency bound for estimating $\E(Y)$ is given by
\begin{align}
\thetaonezeroA:=\E\{(1-\nu)(Y-\beta_{\!\Asc}\trans\,\Xarrow_{\!\!\Asc})^2+\nu(Y-\mu_0)^2\},\,\hbox{where }\beta_{\!\Asc}:=\E(\Xarrow_{\!\!\Asc}\Xarrow_{\!\Asc}\trans)\minusone\E(\Xarrow_{\!\!\Asc}Y).
\label{equ:bound_biased_average}
\end{align}
This is the same efficiency bound for estimating $\E(Y)$ using $\{(Y_i,X_{i\Asc}):\iton\}$ and $N\minusone\suminonetonN X_{i\Asc}$. Leveraging both $\Dsc\IN\equiv\{(Y_i,X_i):\iton\}$ and $\xitilde$, \citet{hu2022paradoxes} proposed an asymptotically unbiased estimator without prior knowledge of $\Asc$, which attains the efficiency bound $\thetaonezeroA$ in \eqref{equ:bound_biased_average}. However, before acquiring the external information $\xitilde$, it is generally impossible to identify $\Asc$ and $\thetaonezeroA$, so inference on utility measures involving $\thetaonezeroA$ cannot be drawn based solely on the internal data $\Dsc\IN$. In contrast, our utility measure $\theta_0$, defined in \eqref{equ:thetazero_response_mean} with $g(X)\equiv\beta_0\trans\Xarrow$, can be inferred using $\Dsc\IN$. It has a meaningful interpretation for $\xitilde$: because $\theta_0=\thetaonezero^{(\Jsc)}/\thetatwozero=\min_{\Asc\in 2^\Jsc}\thetaonezeroA/\thetatwozero$ for the power set $2^\Jsc$ of $\Jsc$, the complement $(1-\theta_0)$ represents the upper bound (over all possible $\Asc$) of efficiency improvement that can be achieved by incorporating $\xitilde$. Therefore, even when $\xitilde$ is a possibly biased summary statistic, the measure $\theta_0$ still characterizes its maximum potential contribution to the estimation, corresponding to the ideal case where $\xitilde$ is unbiased. When the external information takes other forms, the interpretation of $\theta_0$ is similar. In summary, for possibly biased external information, $\theta_0$ is arguably the most informative and meaningful utility measure that can be inferred based solely on internal data.
  
\subsection{Potential extension}
Thus far, we have been concentrated on the potential efficiency benefits of external information, primarily due to the simplicity of its quantification. However, more broadly, incorporating external information can also improve other aspects of inference. For example, in the context of treatment effect estimation based on observational studies, \citet{chakrabortty2022general} demonstrated that including additional individual data of treatment indicators and baseline covariates can enhance robustness compared to estimators relying solely on the internal data. Future research will aim to quantify the robustness benefits of external information in causal inference and other problems involving model misspecification, thereby expanding the scope and applicability of our framework.

\begin{appendices}
\section{Example: Response quantile estimation with individual covariate data}
\label{sec:response_quantile_individual_data}
In this section, let $Z\equiv(Y,X)\in\R\times\R^p$ where $Y$ is a continuous response variable, and let $\mu_0$ denote the $\tau$-quantile of $Y$ for a predetermined constant $\tau\in(0,1)$. We aim to investigate whether incorporating the external information $\{X_i:\inplusonetonN\}$ can improve the estimation of $\mu_0$ using only the internal data $\{(Y_i,X_i):\iton\}$. Recall $\nu\equiv n/(n+N)\in[\,0,1)$. \citet{chakrabortty2022semi} showed that the efficiency bounds for estimating $\mu_0$ in the data-fusion and internal-data-only settings are given, respectively, by
\begin{align}
\{(1-\nu)\E[\{\one(Y<\mu_{0})-\FYX(\mu_0\mid X)\}^2]+\nu\tau(1-\tau)\} /f_Y(\mu_0)^2 ,\ \tau(1-\tau)/f_Y(\mu_0)^2,
\label{equ:efficiency_bounds_response_quantile}
\end{align}
where $\FYX(\cdot\mid X)$ denotes the conditional distribution function of $Y$ given $X$, and $f_Y$ is the density function of $Y$. The utility measure $\theta_0$ defined in \eqref{equ:def_theta0} then takes the form
\begin{align}
\theta_{0}\equiv(1-\nu)\E[\{\one(Y<\mu_0)-\FYX(\mu_0\mid X)\}^2]\,\big/\,\{\tau(1-\tau)\}+\nu.
\label{equ:thetazero_quantile}
\end{align}
Notice that the common term $f_Y(\mu_0)\minustwo$ cancels out from the two efficiency bounds in \eqref{equ:efficiency_bounds_response_quantile} when considering their ratio. We hence define the following for this example:
\begin{align}
\thetaonezero\equiv (1-\nu)\E[\{\one(Y<\mu_0)-\FYX(\mu_0\mid X)\}^2]+\nu\tau(1-\tau),\ \thetatwozero\equiv\tau(1-\tau),
\label{equ:thetaonezero_thetatwozero_quantile}
\end{align}
with a slight abuse of notation. Here $\thetatwozero$ is known since $\tau$ is deterministic. To apply Algorithm \ref{algo:point_estimator}, we first specify the nonparametric efficient influence functions of $\{\thetaonezero,\theta_0\}$ in the following proposition.
\begin{proposition}
\label{prop:EIF_response_quantile}
The nonparametric efficient influence functions of $\thetaonezero\equiv\theta_1(F_{Z})$ in \eqref{equ:thetaonezero_thetatwozero_quantile} and $\theta_{0}\equiv \thetaonezero/\thetatwozero$ in \eqref{equ:thetazero_quantile} are given by:
\begin{align}
&\thetadot_1(F_{Z};\delta_Z-F_{Z})=
(1-\nu)[2\,\E\{\FYX(\mu_{0}\mid X)\fYX(\mu_{0}\mid X)\}/f_Y(\mu_{0})-1]\{\one(Y<\mu_{0})-\tau\}+ \nonumber \\
& \phantom{\thetadot_1(F_{Z};\delta_Z-F_{Z})\equiv\,}(1-\nu)\{\one(Y<\mu_{0})-\FYX(\mu_{0}\mid X)\}^2+\nu\tau(1-\tau)-\thetaonezero, \label{equ:EIF_quantile} \\
&\thetadot(F_{Z};\delta_Z-F_{Z})=\thetadot_1(F_{Z};\delta_Z-F_{Z})/\{\tau(1-\tau)\}, \nonumber
\end{align}
where $\fYX$ denotes the conditional density function of $Y$ given $X$.
\end{proposition} 
  
\subsection{Point estimation}
Based on the results in Proposition \ref{prop:EIF_response_quantile}, we now implement step 2  of Algorithm \ref{algo:point_estimator} to construct a point estimator for $\thetaonezero$ using $\{(Y_i,X_i):\iton\}$. Define $\muhat$ as the empirical $\tau$-quantile of $\{Y_i:\iton\}$, satisfying the condition $|n\minusone\sumin\{\one(Y_i<\muhat)-\tau\}|\leq n\minusone$. For the right hand side of \eqref{equ:EIF_quantile}, when substituting the estimator $\muhat$ for $\mu_0$ and averaging over $\{(Y_i,X_i):\iton\}$, the term involving $n\minusone\sumin\{\one(Y_i<\muhat)-\tau\}$ is asymptotically negligible at the $n\half$ scale. As a result, the point estimator for $\thetaonezero$ is given by
\begin{align}
\thetahat_1\equiv(1-\nu)n\minusone\sumin\{\one(Y_i<\muhat)-\FhatYX(\muhat\mid X_i)\}^2+\nu\tau(1-\tau).
\label{equ:thetahatone_quantile}
\end{align}
For $\{\FhatYX(\muhat\mid X_i):\iton\}$, we again adopt the cross-fitting strategy. For a fixed integer $M\geq 2$, we randomly divide $\Isc\equiv\{1,\ldots,n\}$ into $M$ disjoint subsets $\{\Isc_1,\ldots,\Isc_M\}$ of sizes as even as possible, and set
\begin{align}
\FhatYX(\muhat\mid X_i)\equiv\summM\FhatYXm(\muhat\mid X_i)\one(i\in\Isc_m),
\label{equ:cf_example2}
\end{align}
where $\FhatYXm$ is an estimator for $\FYX$ using the data $\{Z_i:i\in\Isc\backslash\Isc_m\}$. In practice, $\FhatYXm(\muhat\mid\cdot)$ can be constructed by regressing $\{\one(Y_i<\muhat):i\in\Isc\backslash\Isc_m\}$ on $\{X_i:i\in\Isc\backslash\Isc_m\}$, which is computationally easier than recovering the conditional distribution function $\FYX(y\mid\cdot)$ for all $y\in\R$, and allows for the use of flexible regression or machine learning methods.
  
We then transform $\thetahat_1$ from \eqref{equ:thetahatone_quantile} into a point estimator for $\theta_0$ in \eqref{equ:thetazero_quantile}:
\begin{align}
\thetahat\equiv\thetahat_1/\{\tau(1-\tau)\} \equiv(1-\nu) n\minusone\sumin\{\one(Y_i<\muhat)-\FhatYX(\muhat\mid X_i)\}^2/\{\tau(1-\tau)\}+\nu.
\label{equ:thetahat_response_quantile}
\end{align}
In the following, Assumption \ref{ass:response_quantile} imposes a high-level convergence rate condition on the nuisance estimator $\FhatYXm$, which paves the way for establishing the asymptotic properties of $\thetahat$ in Theorem \ref{thm:exm3}.
  
\begin{assumption}\label{ass:response_quantile}
\rm{
For some positive sequences $a_n$ and $r_F\equiv r_F(n)$ satisfying $a_n+r_F\to 0$ and $n\half a_n\to\infty$, the estimator $\FhatYXm$ in \eqref{equ:cf_example2} satisfies
\begin{align*}
\textstyle\supmuB\int\{\FhatYXm(\mu\mid x)-\FYX(\mu\mid x)\}^2dF_X(x)=O_p(r_F^2)\quad(\mtoM), 
\end{align*} 
where $\Bsc(\mu_0,a_n):=\{\mu\in\R:|\mu-\mu_0|\leq a_n\}$.
}
\end{assumption}

\begin{theorem}
\label{thm:exm3}
Suppose Assumption~\ref{ass:response_quantile} holds and $f_{Y}(\mu_{0})>c$ for some constant $c>0$.
Then, the estimator $\thetahat$ in \eqref{equ:thetahat_response_quantile} satisfies
\begin{align}
\thetahat - \theta_{0} = n\minusone\sumin  \thetadot(F_{Z};\deltaZi-F_{Z})+O_{p}(r_{F}^{2}+n\minusone)+o_{p}(n\minushalf),
\label{equ:expansion_quantile}
\end{align}
with $\thetadot(F_{Z};\deltaZi-F_{Z})$ specified in Proposition \ref{prop:EIF_response_quantile}. It follows that:  
\begin{enumerate}[(a)]
\item $\thetahat$ attains the following convergence rate:
\begin{align}
\thetahat-\theta_0
= O_p(n\minushalf + r_{F}^{2}+n\minusone)+o_p(n\minushalf)=O_p(n\minushalf + r_{F}^{2}); \label{equ:rate_mean_quantile}
\end{align}
      
\item if $r_F^2=o(n\minushalf)$ and $\P\{\thetadot(F_{Z};\delta_Z-F_{Z})= 0\}<1$, then $\thetahat$ is nonparametric efficient, that is, $\thetahat-\theta_0=n\minusone\sumin\thetadot(F_{Z};\deltaZi-F_{Z})+o_p(n\minushalf)$;
      
\item if $\P\{\FYX(\mu_0\mid X)=\tau\}=1$, then $\P\{\thetadot(F_{Z};\delta_Z-F_{Z})= 0\}=1$ and the rate in \eqref{equ:rate_mean_quantile} is improved to 
$
\thetahat-\theta_0=O_p(r_F^2+n\minusone)+o_p(n\minushalf).
$
\end{enumerate}
\end{theorem}
\noindent In \eqref{equ:expansion_quantile}, the term $r_F^2$ reflects the errors from estimating the nuisance function $\FYX$, while $O_p(n\minusone)$ and $o_p(n\minushalf)$ correspond to the remainders resulting from applying the delta method and empirical process theory, respectively. These techniques are necessary to handle the nonlinearity in the ratio structure $\thetaonezero/\thetatwozero$ and the indicator function. Additionally, the condition $\P\{\FYX(\mu_0\mid X)=\tau\}=1$ in Theorem \ref{thm:exm3}(c) implies that the utility measure $\theta_0\equiv (1-\nu)\E[\{\one(Y<\mu_0)-\FYX(\mu_0\mid X)\}^2]/\{\tau(1-\tau)\}+\nu=1$, i.e., individual covariate data are useless for estimating the response quantile $\mu_0$. In this case, the influence function $\thetadot(F_{Z};\delta_Z-F_{Z})$ almost surely vanishes, leading to the improved rate for $\thetahat$.
  
\subsection{Confidence interval construction}
As indicated by Theorem \ref{thm:exm3}(c), the first-order term of $(\thetahat-\theta_0)$ vanishes if $\P\{\FYX(\mu_0\mid X)=\tau\}=1$. Due to similar reasoning as in the first paragraph of Section \ref{sec:CI_mean}, this result implies that $\thetahat$ cannot be used to construct confidence intervals for $\theta_0$. Therefore, we need to propose a different estimator to draw valid inference for $\theta_0$.

The first-order term $n\minusone\sumin\thetadot(F_Z;\deltaZi-F_Z)$ in the expansion \eqref{equ:expansion_quantile} consists of errors arising from estimating $\E[\{\one(Y<\mu_{0})-\FYX(\mu_0\mid X)\}^2]$
(as if $\mu_0$ and $\FYX$ were known) and from estimating $\mu_0$ itself, as shown in the Supplementary Material. Notably, these error terms may cancel out. To avoid this, we now approximate $\E[\{\one(Y<\mu_{0})-\FYX(\mu_0\mid X)\}^2]$ and $\mu_0$ using different halves of the available data. This leads to an alternative estimator for $\theta_0$ in \eqref{equ:thetazero_quantile}:
\begin{equation}
\label{equ:def_thetatilde_exm3}
\thetatilde:=(1-\nu)\ntilde\minusone\sumintilde\{\one(Y_i<\mutilde)-\FhatYX(\mutilde\mid X_i)\}^2/\{\tau(1-\tau)\}+\nu   ~\hbox{with}~\ntilde\equiv\ceil{n/2},
\end{equation}
where $\mutilde$ is the empirical $\tau$-quantile of $\{Y_i:\intildeton\}$, and $\{\FhatYX(\mutilde\mid X_i):\itontilde\}$ are constructed using a cross-fitting procedure similar to \eqref{equ:cf_example2}. The first-order term in the expansion of $(\thetatilde-\theta_0)$ is given by $(1-\nu)(\Ttilde_1+\Ttilde_2)/\{\tau(1-\tau)\}$, where
\begin{align*}
&\Ttilde_{1}:=\ntilde\minusone\sumintilde\{\one(Y_i<\mu_{0})-\FYX(\mu_{0}\mid X_i)\}^2-\E[\{\one(Y<\mu_{0})-\FYX(\mu_{0}\mid X)\}^2], \\
&\Ttilde_{2}:=[2\,\E\{\FYX(\mu_{0}\mid X)\fYX(\mu_{0}\mid X)\}/f_Y(\mu_{0})-1](n-\ntilde)\minusone\sumintilden\{\one(Y_i<\mu_{0})-\tau\}.
\end{align*}
Since $\Ttilde_1$ and $\Ttilde_2$ involve different halves of the data, they do not cancel out, ensuring the term $(1-\nu)(\Ttilde_1+\Ttilde_2)/\{\tau(1-\tau)\}$ does not vanishes. We proceed to analyze the limiting distribution of $\thetatilde$ and construct an asymptotic confidence interval for $\theta_0$ in Theorem \ref{thm:exm3_inference}.
  
\begin{theorem}
\label{thm:exm3_inference}
Suppose Assumption~\ref{ass:response_quantile} holds with $r_F^2=o(n\minushalf)$, and there exist some positive constants $\{c_1,c_2\}$ such that $f_{Y}(\mu_{0})>c_1$ and
\begin{align*}
\gamma^{2}:=\ &2(1-\nu)^{2}\,[2\,\E\{\FYX(\mu_{0}\mid X)\fYX(\mu_{0}\mid X)\}/f_Y(\mu_{0})-1]^{2}\,\big/\,\{\tau(1-\tau)\} + \\ &\,2(1-\nu)^{2}\,\Var[\{\one(Y<\mu_0)-\FYX(\mu_0\mid X)\}^2]\,\big/\,\{\tau(1-\tau)\}^2>c_2.
\end{align*}
Then the following results hold:
\begin{enumerate}[(a)]
\item The estimator $\thetatilde$ in \eqref{equ:def_thetatilde_exm3} satisfies $n\half(\thetatilde-\theta_0)/\gamma\,\xrightarrow{w}\,\Normal(0,1)$.
      
\item Let $\{\fhat_Y,\fhatYX\}$ be consistent kernel density estimators for $\{f_Y, \fYX\}$. Define
\begin{align*}
\gammahat^{2}:=\ &2(1-\nu)^{2}\,[2\,n\minusone\sumin\{\FhatYX(\muhat\mid X_i)\fhatYX(\muhat\mid X_i)\}/\fhat_Y(\mu_{0})-1]^{2}\,\big/\,\{\tau(1-\tau)\} + \\ &\,2(1-\nu)^{2}\,\Varhat[\{\one(Y<\muhat)-\FhatYX(\muhat\mid X)\}^2]\,\big/\,\{\tau(1-\tau)\}^2,
\end{align*}
with $\muhat$ the empirical $\tau$-quantile $\{Y_i:\iton\}$ and $\FhatYX$ as in \eqref{equ:cf_example2}. Then, for any predetermined confidence level $\alpha\in(0,1)$, the interval
\begin{align}
\CI_\alpha\!:=\![\,\thetatilde-\gammahat\,u_{(1+\alpha)/2}/n\half,\, \thetatilde+\gammahat\,u_{(1+\alpha)/2}/n\half\,]\hbox{ satisfies }\limninfty\,\P(\theta_0\in\CI_\alpha)=\alpha.
\label{equ:CI_quantile}
\end{align}
\end{enumerate}
\end{theorem}

\section{Example: Multiple linear regression with univariate regression parameter estimates}\label{sec:linear_regression}
\paragraph{Notations.} The notation $\Diag(a_1,\ldots,a_p)$ denotes a diagonal matrix with diagonal entries $a_1,\ldots,a_p$, while $\Tr(\cdot)$ represents the trace of a square matrix. Additionally, let $\Ehat(\cdot)$ denote the sample average computed from $\{Z_i:\iton\}$. 
    
In this section, suppose that $Z\equiv(Y,X)\equiv(Y,S,W)\in\R\times\R\times\R$ follows the model $Y=\mu_0\trans X+\ve$ for a parameter vector $\mu_0\in\R^2$, where the covariate vector $X$ has mean zero, and the random error $\ve$ satisfies $\E(\ve\mid X)=0$ and $\E(\ve^2\mid X)=\sigma_0^2$ for some constant $\sigma_0\in(0,\infty)$. Our goal is to estimate the regression parameter vector $\mu_0$. Beyond the internal data $\{(Y_i,S_i,W_i):\iton\}$, we may have access to additional external information, specifically the quantity $\etatilde:=(\suminonetonN S_iY_i)/(\suminonetonN S_i^2)$. This represents the ordinary least squares estimate based on $N$ additional independent copies $\{(Y_i,S_i):\inplusonetonN\}$ of $(Y,S)$ and targets the parameter $\eta_0:=\E(S\,Y)/\E(S^2)$, which corresponds to a working linear model between $Y$ and $S$. Such a data structure is commonly encountered in genome-wide association studies, where researchers report summary statistics of univariate regression coefficients relating a quantitative trait ($Y$) to single centered genotypes ($S$) \citep{zhu2017bayesian}. It is noteworthy that $\eta_0$ is not necessarily a component of $\mu_0$. 
  
For estimating the target $\mu_0$, \citet{hu2022paradoxes} established that the efficiency bounds in the data-fusion and internal-data-only settings are given, respectively, by $\Theta\DF\equiv\sigma_0^2\,\Sigma_0\minusone-(1-\nu)\,\Diag(\sigma_0^4/\alpha_0,0)$ and $\Theta\IN\equiv\sigma_0^2\Sigma_0\minusone$,
where $\Sigma_0:=\E(XX\trans)$ and $\alpha_0:=\E\{S^2(Y-\eta_0 S)^2\}$.
The parameters $\{\thetaonezero,\thetatwozero\}$ emerge as the traces of these matrices:
\begin{align}
\thetaonezero\equiv\Tr(\Theta\DF)=\sigma_0^2\,\kappa_0-(1-\nu)\sigma_0^4/\alpha_0,\ \thetatwozero\equiv\Tr(\Theta\IN)=\sigma_0^2\,\kappa_0, \hbox{ with } \kappa_0:=\Tr(\Sigma_0\minusone). 
\label{equ:theta01_theta02_linear_regression}
\end{align}
This leads to the utility measure for the external information $\etatilde$, given by
\begin{align}
\theta_{0}\equiv\thetaonezero/\thetatwozero=1-(1-\nu)\sigma_0^2/(\alpha_0\kappa_0).
\label{equ:thetazero_linear_regression}
\end{align}
The nonparametric efficient influence functions of $\{\thetaonezero,\thetatwozero,\theta_0\}$ are specified as follows.
\begin{proposition}\label{prop:EIF_linear_regression}
The nonparametric efficient influence functions of $\{\thetajzero\equiv\theta_j(F_Z):j=1,2\}$ in \eqref{equ:theta01_theta02_linear_regression} and $\theta_0\equiv \thetaonezero/\thetatwozero$ in \eqref{equ:thetazero_linear_regression} are given by:
\begin{align*}
&\thetadot_1(F_{Z};\delta_Z-F_{Z}) =\kappa_0\ve^2-\sigma_0^2\{\Tr(\Sigma_0\minusone XX\trans\Sigma_0\minusone)-\kappa_0\}-2(1-\nu)(\sigma_0^2/\alpha_0)\ve^2+(1-\nu)(\sigma_0^4/\alpha_0^2)\\
&\phantom{\thetadot_1(F_{Z};\delta_Z-F_{Z}) =}\times[S^2(Y-\eta_0 S)^2-2\,\E\{S^3(Y-\eta_0 S)\}S(Y-\eta_0S)/\E(S^2)]-\thetaonezero, \\
&\thetadot_2(F_{Z};\delta_Z-F_{Z})=\kappa_0\ve^2-\sigma_0^2\{\Tr(\Sigma_0\minusone XX\trans\Sigma_0\minusone)-\kappa_0\}-\thetatwozero, \\
&\thetadot(F_{Z};\delta_Z-F_{Z})=\thetadot_1(F_{Z};\delta_Z-F_{Z})/\thetatwozero-\thetaonezero\thetadot_2(F_{Z};\delta_Z-F_{Z})/\thetatwozero^2.
\end{align*}
\end{proposition}
  
\subsection{Point estimation}
Using the internal data $\{(Y_i,S_i,W_i):\iton\}$, we now construct a point estimator for the utility measure $\theta_0$ in \eqref{equ:thetazero_linear_regression}. The ordinary least squares estimators for $\mu_0\equiv\E(XX\trans)\minusone\E(XY)$ and $\eta_0\equiv\E(SY)/\E(S^2)$ are defined as
\begin{align}
\muhat:=(\sumin X_iX_i\trans)\minusone\sumin X_iY_i,\ \etahat:=(\sumin S_iY_i)/(\sumin S_i^2).
\label{equ:muhat_etahat}
\end{align}
Let $\Sigmahat:=n\minusone\sumin X_iX_i\trans$. We then define the estimators for $\kappa_0\equiv\Tr\{\E(XX\trans)\minusone\}$, $\sigma_0\equiv\{\E(\ve^2)\}\half$ and $\alpha_0\equiv\E\{S^2(Y-\eta_0 S)^2\}$ as
\begin{align}
\kappahat:=\Tr(\Sigmahat\minusone),\,  \sigmahat:=\{n\minusone\sumin(Y_i-\muhat\trans X_{i})^2\}\half,\,  \alphahat:=n\minusone\sumin S_i^2(Y_i-\etahat S_i)^2.
\label{equ:kappahat_sigmahat_alphahat}
\end{align}
By substituting these estimates into the influence functions in Proposition \ref{prop:EIF_linear_regression} and averaging over $\{(Y_i,S_i,W_i):\iton\}$, we obtain estimators for $\{\thetaonezero,\thetatwozero\}$:
\begin{align*}
\thetahat_1\equiv\sigmahat^2\kappahat-(1-\nu)\sigmahat^4/\alphahat~\hbox{ and }~ \thetahat_2\equiv\sigmahat^2\kappahat.
\end{align*}
Their ratio provides a point estimator for $\theta_0\equiv\thetaonezero/\thetatwozero$:
\begin{equation}
\thetahat \equiv\thetahat_1\,\big/\,\thetahat_2= 1-(1-\nu)\, \sigmahat^2\,\big/\,(\alphahat\,\kappahat). \label{equ:thetahat_exm4}
\end{equation}
The following theorem establishes that under a mild moment condition, $\thetahat$ attains the influence function $\thetadot(F_Z;\delta_Z-F_Z)$ derived in Proposition \ref{prop:EIF_linear_regression}.
\begin{theorem}
\label{thm:exm4}
Suppose $\E\{\thetadot(F_Z;\delta_Z-F_Z)^2\}<\infty$. Then the estimator $\thetahat$ in \eqref{equ:thetahat_exm4} satisfies
$
\thetahat - \theta_{0} = n\minusone\sumin  \thetadot(F_{Z};\deltaZi-F_{Z})
+ O_{p}(n\minusone) = O_{p}(n\minushalf).
$
\end{theorem}
  
\subsection{Confidence interval construction}\label{sec:CI_linear_regression}
The asymptotic variance $\gamma^2:=\E\{\thetadot(F_Z;\delta_Z-F_Z)^2\}$ of the point estimator $\thetahat$ in \eqref{equ:thetahat_exm4} is positive under reasonable conditions, as demonstrated in Section \ref{sec:supp_CI_linear_regression} of the Supplementary Material. In particular, a sufficient set of conditions for $\gamma^2>0$ is: (i) $\P(S^2\neq\alpha_0/\sigma_0^2)>0$; (ii) $\ve$ is independent of $X$; (iii) the support of $\ve$ contains more than two elements. In the current model, where $\E(\ve\mid X)=0$ and $\E(\ve^2\mid X)=\sigma_0^2$ for some constant $\sigma_0\in(0,\infty)$, conditions (ii)--(iii) hold, for example, when $\ve$ follows a normal distribution. Consequently, we can directly use $\thetahat$ to construct confidence intervals for $\theta_0$ without requiring an alternative estimator, as was necessary in the examples from Section \ref{sec:mean_estimation} and Appendix \ref{sec:response_quantile_individual_data}.
  
\begin{theorem}\label{thm:CI_linear_regression}
Suppose $\gamma^2\equiv\E\{\thetadot(F_Z;\delta_Z-F_Z)^2\}\in(0,\infty)$. Then, the asymptotic distribution of $\thetahat$ in \eqref{equ:thetahat_exm4} is $n\half(\thetahat-\theta_0)/\gamma\xrightarrow{w}\Normal(0,1)$. Furthermore, define 
\begin{align*}
&\gammahat^2:=\Big (\frac{1-\nu}{\alphahat\,\kappahat}\Big)^2\,\Varhat\Big(
(Y-\muhat\trans X)^2+ (\sigmahat^2/\kappahat)\Tr(\hat{\Sigma}\minusone XX\trans\hat{\Sigma}\minusone)-  \\
& \phantom{\gammahat^2:=\Big (\frac{1-\nu}{\alphahat\,\kappahat}\Big)^2\,\Varhat\Big(
\,} (\sigmahat^2/\alphahat) [S^2(Y-\etahat S)^2-2\,\Ehat\{S^3(Y-\etahat S)\}S(Y-\etahat S)/\Ehat(S^2)]\Big),
\end{align*}
where $\Sigmahat\equiv n\minusone\sumin X_iX_i\trans$ and $\{\muhat,\etahat,\kappahat,\sigmahat,\alphahat\}$ are defined as in \eqref{equ:muhat_etahat}--\eqref{equ:kappahat_sigmahat_alphahat}. Then, for any predetermined confidence level $\alpha\in(0,1)$, the interval
\begin{align}
\CI_\alpha\!:=\![\,\thetahat-\gammahat\,u_{(1+\alpha)/2}/n\half,\, \thetahat+\gammahat\,u_{(1+\alpha)/2}/n\half\,]\,\hbox{ satisfies }\,\limninfty\,\P(\theta_0\in\CI_\alpha)=\alpha.
\label{equ:CI_linear_regression}
\end{align}
\end{theorem}

\section{Additional numerical results}\label{sec:additional_numerical_results}
In the setting of Section \ref{sec:simu}, we now conduct simulation studies for the quantile estimation and linear regression examples in Appendices \ref{sec:response_quantile_individual_data} and \ref{sec:linear_regression}, where the true values of the utility measure $\theta_0$ are given by
\begin{align*}
\left\{ 
\begin{aligned}
& (1-\nu)\{\tau-\E(\Phi[u_\tau\{2b^{2}(1+\rho)+1\}\half -b(S+W)]^2)\}/\{\tau(1-\tau)\} + \nu,\ \hbox{Appendix \ref{sec:response_quantile_individual_data}}; \\
& 1-(1-\nu)(1-\rho^2)\,\big/\,[2\{1+b^{2}(1-\rho^{2})\}],\ \hbox{Appendix \ref{sec:linear_regression}}. 
\end{aligned}
\right.
\end{align*}
For the quantile estimation example, we consider two quantile levels, $\tau\in\{0.25,0.50\}$. To construct $\FhatYXm(\mu\mid\cdot)$, the estimator for the conditional distribution function $\FYX(\mu\mid\cdot)$ with $\mu\in\{\muhat,\mutilde\}$, we perform local linear regression of $\{\one(Y_i<\mu):i\in\Isc\backslash\Isc_m\}$ on $\{X_i:i\in\Isc\backslash\Isc_m\}$. The marginal and conditional density functions, $f_Y$ and $\fYX$, are approximated via kernel smoothing. All bandwidths are selected using the rule of thumb \citep{silverman2018density}.

\begin{table}
\begin{center}
\caption{Simulation results for the point estimator $\thetahat$ in \eqref{equ:thetahat_response_quantile} and $95\%$ confidence interval $\CI_{0.95}$ in \eqref{equ:CI_quantile}: Means (MAE) and standard deviations (SDAE) of the absolute error $|\thetahat-\theta_0|$, as well as the average lengths (AL) and coverage rates (CR) of $\CI_{0.95}$. Here $b$ is the coefficient in model \eqref{equ:simulation_model}, $n$ is the sample size of internal data, and $\tau$ is the quantile level. \ul{The numbers in all columns, except for the first and second, have been multiplied by $100$}. \label{tab:simulation_response_quantile}
\vspace{3mm}}
\begin{tabular}{|cc|cccc|cccc|}
\hline
&      & \multicolumn{4}{c|}{$\tau=0.25$} & \multicolumn{4}{c|}{$\tau=0.50$} \\
$b$                    & $n$  & MAE    & SDAE   & AL     & CR   & MAE    & SDAE   & AL     & CR   \\
\hline
\multirow{3}{*}{$0$}   & 500  & 0.0150 & 0.0786 & 11.9521 & 97.4 & 0.0054 & 0.0403 & 2.1161  & 99.7 \\
& 1000 & 0.0052 & 0.0338 & 8.4629  & 97.6 & 0.0046 & 0.0366 & 1.0689  & 99.6 \\
& 2000 & 0.0041 & 0.0236 & 5.9291  & 97.8 & 0.0019 & 0.0178 & 0.5427  & 99.6 \\
\hline
\multirow{3}{*}{$0.5$} & 500  & 1.9798 & 1.3743 & 15.4597 & 93.0 & 1.8453 & 1.3238 & 9.6217  & 91.1 \\
& 1000 & 1.2433 & 0.8844 & 11.3918 & 93.5 & 1.1320 & 0.7911 & 6.5765  & 95.0 \\
& 2000 & 0.8230 & 0.6003 & 8.0750  & 94.8 & 0.7351 & 0.5341 & 4.5623  & 93.6 \\
\hline
\multirow{3}{*}{$1.0$} & 500  & 2.5285 & 1.6565 & 14.6512 & 92.5 & 2.1337 & 1.4778 & 10.0685 & 90.4 \\
& 1000 & 1.5691 & 1.0946 & 10.0571 & 93.0 & 1.2930 & 0.8966 & 6.7997  & 93.0 \\
& 2000 & 1.1746 & 0.7784 & 7.0493  & 92.9 & 0.8860 & 0.6259 & 4.6828  & 90.7  \\
\hline
\end{tabular}
\end{center}
\end{table}

\begin{table}
\begin{center}
\caption{Simulation results for the point estimator $\thetahat$ in \eqref{equ:thetahat_exm4} and $95\%$ confidence interval $\CI_{0.95}$ in \eqref{equ:CI_linear_regression}: Means (MAE) and standard deviations (SDAE) of the absolute error $|\thetahat-\theta_0|$, as well as the average lengths (AL) and coverage rates (CR) of $\CI_{0.95}$. Here $b$ is the coefficient in model \eqref{equ:simulation_model}, and $n$ is the sample size of internal data. \ul{The numbers in all columns, except for the first, have been multiplied by $100$}. \label{tab:simulation_linear_regression} \vspace{3mm}}
\resizebox{\textwidth}{!}{
\begin{tabular}{|c|cccc|cccc|cccc|}
\hline
& \multicolumn{4}{c|}{$n=500$}     & \multicolumn{4}{c|}{$n=1000$}    & \multicolumn{4}{c|}{$n=2000$}    \\
$b$   & MAE    & SDAE   & AL     & CR   & MAE    & SDAE   & AL     & CR   & MAE    & SDAE   & AL     & CR   \\
\hline
$0$   & 1.9949 & 1.4742 & 9.0991 & 92.9 & 1.3243 & 1.0362 & 6.5792 & 94.9 & 0.9951 & 0.7432 & 4.7075 & 93.5 \\
$0.5$ & 1.6189 & 1.1954 & 7.6129 & 94.4 & 1.1183 & 0.8633 & 5.4810 & 94.4 & 0.8315 & 0.6281 & 3.9344 & 93.7 \\
$1.0$ & 1.0382 & 0.7718 & 5.0699 & 94.8 & 0.7471 & 0.5792 & 3.6447 & 94.7 & 0.5510 & 0.4057 & 2.6176 & 94.7 \\
\hline
\end{tabular}}
\end{center}
\end{table}

Tables \ref{tab:simulation_response_quantile} and \ref{tab:simulation_linear_regression} present the results for the quantile estimation and linear regression examples, respectively. Our method performs well, yielding point estimates with negligible errors and confidence intervals with satisfactory coverage rates. Notably, the over-coverage observed in Table \ref{tab:simulation_response_quantile} when $b=0$ stems from the truncation operation \eqref{equ:truncation}; see the second paragraph of Section \ref{sec:simu} for detailed explanations. In summary, these simulations validate the theoretical results in Appendices \ref{sec:response_quantile_individual_data} and \ref{sec:linear_regression}, further confirming our method’s ability to draw valid inference on the utility of external information.
\end{appendices}

  
\section*{Funding}
Guorong Dai was supported by National Natural Science Foundation of China (12401342 and 72271060). 
Lingxuan Shao was supported by National Natural Science Foundation of China (12401340 and 12471279) and Shanghai Committee of Science and Technology (24ZR1405700).
Jinbo Chen was supported by National Institute of Health (R01-HL138306, R01-CA236468 and R01-GM140463).

\section*{Data availability}
All programs and data used in the numerical studies are available at \url{https://github.com/LingxuanShao/UDF}.


\section*{Supplementary materials}
Necessary supplements to the main article, as well as all technical proofs, can be found in the Supplementary Material.
  
\bibliographystyle{plainnat}
\bibliography{UDF}

\end{document}